\documentclass{my-elsarticle}

\usepackage{a4wide}
\usepackage{amssymb,amsmath,amsthm}
\usepackage{epsfig,graphicx,epstopdf}
\usepackage{hyperref,color}

\allowdisplaybreaks

\theoremstyle{definition}

\begin{document}
	
\begin{frontmatter}
		
\title{\text{Fractional model of COVID-19 applied to Galicia, Spain and Portugal}}

\author[Add:a,Add:b]{\href{https://orcid.org/0000-0002-0119-6178}{Fa\"{\i}\c{c}al Nda\"{\i}rou}} 
\ead{faical@ua.pt}

\author[Add:b]{\href{https://orcid.org/0000-0003-0872-5017}{Iv\'an Area}}
\ead{area@uvigo.es}

\author[Add:c]{\href{https://orcid.org/0000-0001-8202-6578}{Juan J. Nieto}}
\ead{juanjose.nieto.roig@usc.es}

\author[Add:a]{\href{https://orcid.org/0000-0002-7238-546X}{Cristiana J. Silva}} 
\ead{cjoaosilva@ua.pt}

\author[Add:a]{\href{https://orcid.org/0000-0001-8641-2505}{Delfim F. M. Torres\corref{corD}}} 
\ead{delfim@ua.pt}
\cortext[corD]{Corresponding author: Delfim F. M. Torres (delfim@ua.pt)}

\address[Add:a]{Center for Research and Development in Mathematics and Applications (CIDMA),\\ 
Department of Mathematics, University of Aveiro, 3810-193 Aveiro, Portugal}

\address[Add:b]{Departamento de Matem\'atica Aplicada II, 
E. E. Aeron\'autica e do Espazo, Campus de Ourense,\\ 
Universidade de Vigo, 32004 Ourense, Spain} 

\address[Add:c]{Instituto de Matematicas, Universidade de Santiago de Compostela, 
15782 Santiago de Compostela, Spain}


\begin{abstract}
A fractional compartmental mathematical model for the spread of the COVID-19 
disease is proposed. Special focus has been done on the transmissibility 
of super-spreaders individuals. Numerical simulations are shown for data 
of Galicia, Spain, and Portugal. For each region, the order of the Caputo 
derivative takes a different value, that is not close to one, 
showing the relevance of considering fractional models. 
\end{abstract}

\begin{keyword}
mathematical modeling of COVID-19 pandemic 
\sep Galicia, Spain and Portugal case studies
\sep fractional differential equations
\sep numerical simulations.

\medskip

\MSC[2010]{26A33 \sep 34A08 \sep 92D30.}
\end{keyword}

\end{frontmatter}


\section{Introduction}

Coronavirus disease 2019 (COVID-19), the outbreak due to severe acute 
respiratory syndrome coronavirus 2 (SARS-CoV-2), has taken on pandemic 
proportions in 2020, affecting several millions of individuals in almost 
all countries \cite{Hopkins}. An integrated science and multidisciplinary 
approach is necessary to fight the COVID-19 pandemic \cite{Mohamed,Moradian}. 
In particular, mathematical and epidemiological simulation plays a crucial 
role in predicting, anticipating, and controlling present and future epidemics. 

As for the mathematical modelling of coronavirus disease COVID-19, 
it has been shown to be extremely useful for governments in order 
to define appropriate policies \cite{FANT}. In this direction, 
a number of papers has been recently published related 
with modelling of this pandemic (see, e.g., \cite{Gatto,Giordano}, 
just to cite some of them). 

In \cite{FANT}, a model including the 
super-spreader class has been presented, and applied to give an estimation 
of the infected and death individuals in Wuhan. The collaboration with 
Galician government \cite{ANHN} has allowed to understand some important 
considerations in order to perform analysis. In particular, due to the pandemic, 
some cases have not been reported as expected, but with some days of delay. 
As a consequence, in this paper we propose to consider not the daily reported cases, 
but the means in the previous 5 days of daily reported cases. As a result, 
it seems appropriate to consider fractional derivatives, which have been intensively 
used to obtain models of infectious diseases since they take into account the memory effect, 
which is now bigger due to the aforementioned mean of the five previous days of daily reported cases. 
Having estimates {\em a priori} of infected individuals of COVID-19, obtained by using mathematical models, 
has helped to predict the number of required beds both for hospitalized individuals 
and mainly at intensive care units \cite{ANHN}.

Fractional calculus and fractional differential equations have recently been applied 
in numerous areas of mathematics, physics, engineering, bio-engineering, 
and other applied sciences. We refer the reader to the monographs 
\cite{Ge_Chen_Kou,H,KiST,SaKM,T,VC,Zh} and the articles
\cite{Agarwal_Baleanu_Nieto_Torres_Zhou,Alshabanat_Jleli_Kumar_Samet,Nisar,Yildiz}. 
In this work we shall consider the Caputo fractional derivative \cite{Caputo} 
(see also \cite{Gerasimov}). A fractional model using the Caputo--Fabrizio fractional 
derivative of COVID-19 in Wuhan (China) has been developed in \cite{Ram}.

The structure of this work is as follows. In Section~\ref{sec:model}, 
we introduce a fractional model by using Caputo fractional derivatives 
on the classical compartmental model presented in \cite{FANT}, and
where the fractional order of differentiation $\alpha$ can be used 
to describe different strains and genomes of the coronavirus 
and vary with mutations. In Section~\ref{sec:numerical}, 
some numerical results are presented for three 
different territories: Galicia, Spain, and Portugal. Galicia 
is an autonomous community of Spain and located in the northwest Iberian 
Peninsula and having a population of about 2,700,000 and a total area of 29,574 km2. 
Spain (officially, the Kingdom of Spain) is a country mostly located on the Iberian Peninsula, 
in southwestern Europe, with a population of about 47,000,000 people and a total area of 
505,992 km2. Portugal (officially, the Portuguese Republic) is also a country located mostly 
on the Iberian Peninsula with a population of about 10,276,000 individuals 
and a total area of 92,212 km2. We end with Section~\ref{sec:conc} 
of conclusions and discussion. 


\section{The Proposed COVID-19 Fractional Model}
\label{sec:model}

In what follows we shall assume that we have a constant population 
divided in 8 epidemiological classes, namely:
\begin{enumerate} 
\item susceptible individuals ($S$), 
\item exposed individuals ($E$), 
\item symptomatic and infectious individuals ($I$), 
\item super-spreaders individuals ($P$),
\item infectious but asymptomatic individuals ($A$), 
\item hospitalized individuals ($H$), 
\item recovery individuals ($R$), and
\item dead individuals ($F$) or fatality class.
\end{enumerate}

Our model is based on the one presented in \cite{FANT} and substituting 
the first order derivative by a derivative of fractional order $\alpha$. We use 
the fractional derivative in the sense of Caputo: for an absolutely continuous 
function $f: [0, \infty) \to \mathbb{R}$ the Caputo fractional derivative 
of order $\alpha > 0$ is given by \cite{H,KiST,Kumar3,SaKM}: 
\[
D^{\alpha} f(t)=\frac{1}{\Gamma(1-\alpha)} \int_{0}^{t} (t-s)^{-\alpha} f'(s)ds.
\]

\begin{figure}[h!]
\label{figure:2}
\centering
\includegraphics[scale=0.4]{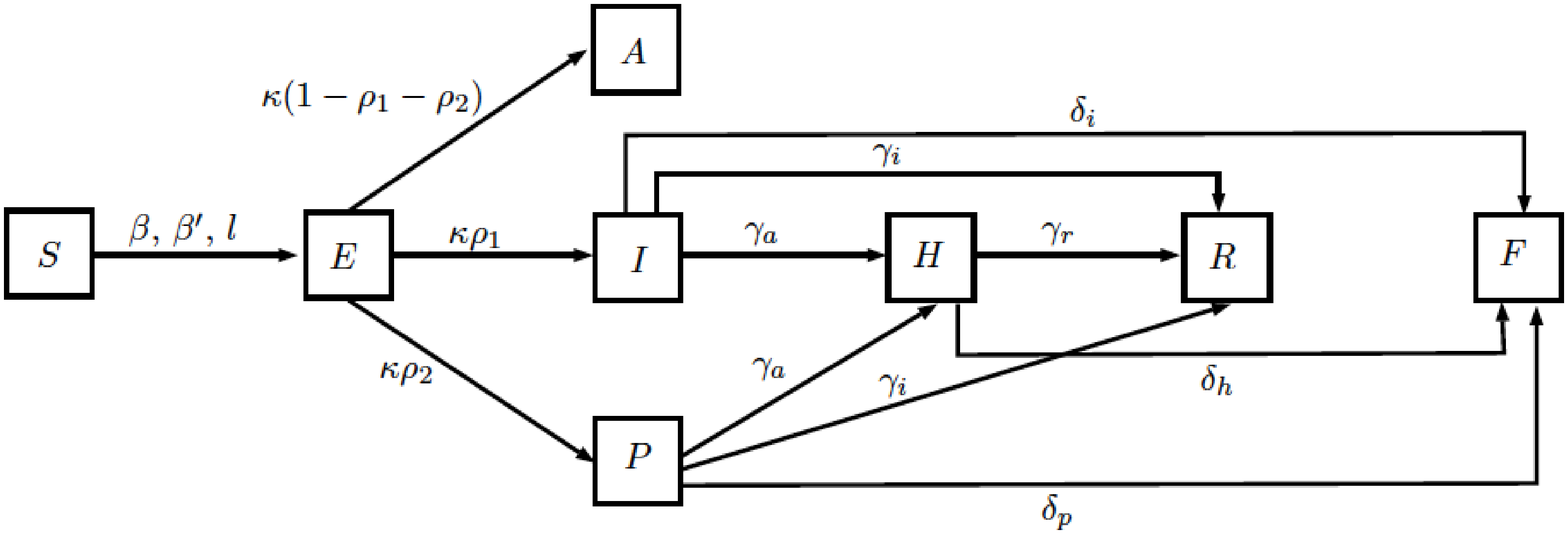}
\caption{Flowchart of model \eqref{model}.}
\end{figure}

Fractional calculus and fractional differential equations are an active area 
of research and, in some cases, adequate to incorporate the history of the processes \cite{Agarwal_Baleanu_Nieto_Torres_Zhou,Goswami1,Kumar1,Kumar3,Kumar2,Singh1,Zh}. 
The fractional proposed model takes the form
\begin{equation}
\label{model}
\begin{cases}
\displaystyle{ D^{\alpha}S(t) 
= -\beta\frac{I}{N}S-l\beta \frac{H}{N}S-\beta^{'}\frac{P}{N}S},\\[2mm]
\displaystyle{ D^{\alpha}E(t)
= \beta\frac{I}{N}S+l\beta \frac{H}{N}S+ \beta^{'}\frac{P}{N}S -\kappa E}, \\[2mm]
\displaystyle{ D^{\alpha}I(t)
= \kappa \rho_1 E - (\gamma_a + \gamma_i)I-\delta_i I}, \\[2mm]
\displaystyle{ D^{\alpha}P(t)
= \kappa \rho_2 E- (\gamma_a + \gamma_i)P-\delta_p P}, \\[2mm]
\displaystyle{ D^{\alpha}A(t)
= \kappa (1-\rho_1 - \rho_2)E },\\[2mm]
\displaystyle{ D^{\alpha}H(t)
= \gamma_a (I + P) - \gamma_r H - \delta_h H}, \\[2mm]
\displaystyle{ D^{\alpha}R(t)
= \gamma_i (I + P)+ \gamma_r H},\\[2mm]
\displaystyle{ D^{\alpha}F(t)
= \delta_i I(t) + \delta_p P(t) + \delta_h H(t)},
\end{cases}
\end{equation}
in which we have the following parameters:
\begin{enumerate}
\item $\beta$ quantifies the human-to-human transmission coefficient per unit time (days) per person, 
\item $\beta^{'}$ quantifies a high transmission coefficient due to super-spreaders, 
\item $l$ quantifies the relative transmissibility of hospitalized patients,
\item $\kappa$ is the rate at which an individual leaves the exposed 
class by becoming infectious (symptomatic, super-spreaders or asymptomatic),
\item $\rho_1$ is the proportion of progression from exposed class $E$ to symptomatic infectious class $I$,
\item $\rho_2$ is a relative very low rate at which exposed individuals become super-spreaders,
\item $1-\rho_1-\rho_2$ is the progression from exposed to asymptomatic class,
\item $\gamma_a$ is the average rate at which symptomatic  and super-spreaders individuals become hospitalized,
\item $\gamma_i$ is the recovery rate without being hospitalized,
\item $\gamma_r$ is the recovery rate of hospitalized patients,
\item $\delta_i$ denotes the disease induced death rates due to infected individuals, 
\item $\delta_p$ denotes the disease induced death rates due to super-spreaders individuals,
\item $\delta_h$ denotes the disease induced death rates due to hospitalized individuals.
\end{enumerate}
A flowchart of model \eqref{model} is presented 
in Figure~\ref{figure:2}. For additional details and particular 
values of the parameters we refer the reader to \cite{FANT}.

\section{Numerical Simulations}
\label{sec:numerical}

Next, we shall show the numerical simulations in three territories: Galicia, Portugal, and Spain. 
For all these cases we have considered the official data published by the corresponding 
authorities and we have computed the means of the five previous reports. As it has been 
observed during this pandemic, the output of the laboratories has had some delays due 
to the big pressure and collapse of the public health systems. In this way, some cases have been 
reported with some delay and some updates have been published days later of the corresponding dates. 
In order to reduce these problems, we consider the mean of the five previous reported cases, 
always following the official data. Moreover, in each of the territories there 
are specificities such as territorial dispersion/concentration, use of public transportation, 
and mainly the date of starting the confinement, as compared with the initial spread of the COVID-19. 
These factors imply tiny adjustments in the factor to divide the total population as well as in 
the value of the fractional parameter $\alpha$. For solving the system of fractional 
differential equations \eqref{model} we have used \cite{DIETHELM}, by using Matlab in a MacBook Pro 
computer with a 2.3 GHz Intel Core i9 processor and 16GB of 2400 MHz DDR4 memory.


\subsection{The Case Study of Galicia}

In the autonomous region of Galicia, we have the values given in 
Table~\ref{table:galiza} as for the cumulative cases, 
the new daily infected individuals, as well as the mean of the $5$ previous days. 
\begin{table}[ht!] 
\label{table:galiza}
\begin{tabular}{|l|l|l|l|l|l|l|l|l|} \hline
Date & Confirmed & \begin{tabular}{l} New \\ 
confirmed \end{tabular} & \begin{tabular}{l} 5 days \\ mean \end{tabular} & & Date & Confirmed & 
\begin{tabular}{l} 
New \\ confirmed 
\end{tabular}  &  
\begin{tabular}{l} 5 days \\ mean 
\end{tabular}  \\ \hline
03-08 & 6    & 1   & 1     &  &  04-03 & 5625 & 406 & 380,4 \\ \hline
03-09 & 22   & 16  & 4,2   &  &  04-04 & 5944 & 319 & 381   \\ \hline
03-10 & 35   & 13  & 6,4   &  &  04-05 & 6151 & 207 & 343,8 \\ \hline
03-11 & 35   & 0   & 6,4   &  &  04-06 & 6331 & 180 & 297,8 \\ \hline
03-12 & 85   & 50  & 16    &  &  04-07 & 6538 & 207 & 263,8 \\ \hline
03-13 & 115  & 30  & 21,8  &  &  04-08 & 6758 & 220 & 226,6 \\ \hline
03-14 & 195  & 80  & 34,6  &  &  04-09 & 6946 & 188 & 200,4 \\ \hline
03-15 & 245  & 50  & 42    &  &  04-10 & 7176 & 230 & 205   \\ \hline
03-16 & 292  & 47  & 51,4  &  &  04-11 & 7336 & 160 & 201   \\ \hline
03-17 & 341  & 49  & 51,2  &  &  04-12 & 7494 & 158 & 191,2 \\ \hline
03-18 & 453  & 112 & 67,6  &  &  04-13 & 7597 & 103 & 167,8 \\ \hline
03-19 & 578  & 125 & 76,6  &  &  04-14 & 7708 & 111 & 152,4 \\ \hline
03-20 & 739  & 161 & 98,8  &  &  04-15 & 7873 & 165 & 139,4 \\ \hline
03-21 & 915  & 176 & 124,6 &  &  04-16 & 8013 & 140 & 135,4 \\ \hline
03-22 & 1208 & 293 & 173,4 &  &  04-17 & 8084 & 71  & 118   \\ \hline
03-23 & 1415 & 207 & 192,4 &  &  04-18 & 8185 & 101 & 117,6 \\ \hline
03-24 & 1653 & 238 & 215   &  &  04-19 & 8299 & 114 & 118,2 \\ \hline
03-25 & 1915 & 262 & 235,2 &  &  04-20 & 8468 & 169 & 119   \\ \hline
03-26 & 2322 & 407 & 281,4 &  &  04-21 & 8634 & 166 & 124,2 \\ \hline
03-27 & 2772 & 450 & 312,8 &  &  04-22 & 8805 & 171 & 144,2 \\ \hline
03-28 & 3139 & 367 & 344,8 &  &  04-23 & 8932 & 127 & 149,4 \\ \hline
03-29 & 3723 & 584 & 414   &  &  04-24 & 9116 & 184 & 163,4 \\ \hline
03-30 & 4039 & 316 & 424,8 &  &  04-25 & 9176 & 60  & 141,6 \\ \hline
03-31 & 4432 & 393 & 422   &  &  04-26 & 9238 & 62  & 120,8 \\ \hline
04-01 & 4842 & 410 & 414   &  &  04-27 & 9328 & 90  & 104,6 \\ \hline
04-02 & 5219 & 377 & 416   &  &            &      &     &      \\ \hline
\end{tabular} 
\caption{Data of the autonomous region of Galicia. The list of 51 days 
includes the cumulative, new infected and mean of the previous $5$ days.}
\end{table}

The data includes 51 values starting 7th March since after that date 
(27th April) the way of officially computing individuals has changed.

\begin{figure}[!htb]
\centering \label{figure:galiza}
\includegraphics[scale=0.4]{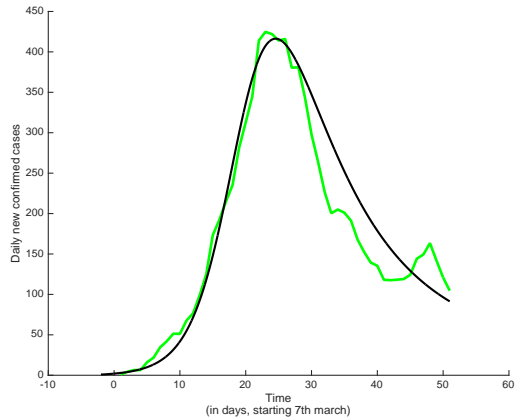}
\caption{Number of confirmed cases per day in Galicia. The green line corresponds 
to the real data given in Table~\ref{table:galiza} while the black line ($I + P + H$) 
has been obtained by solving numerically the system of fractional differential 
equations \eqref{model}, by using \cite{DIETHELM}.}
\end{figure}

By considering the fractional order $\alpha=0.85$ and the same values 
of the parameters as in \cite{FANT}, the results of the numerical 
simulation are shown in Figure~\ref{figure:galiza}. 
The green line denotes the real data while the 
black line is the numerical solution of the fractional system \eqref{model}, 
with total population $N=2,700,000/500$, where $N = S + E + I + P + A + H + R + F$, 
since the population of Galicia is widely dispersed in the territory with very 
few big cities and low use of public transportation.


\subsection{The Case Study of Spain}

As for the Kingdom of Spain, the data of 82 days is collected in 
Table~\ref{table:spain}, as for the cumulative cases, 
the new daily infected individuals, as well as the mean 
of the $5$ previous days, starting 25th February.
\begin{table}[ht!] 
\label{table:spain}
\begin{tabular}{|l|l|l|l|l|l|l|l|l|} \hline
Date & Confirmed & 
\begin{tabular}{l} 
New \\ 
confirmed 
\end{tabular} 
& \begin{tabular}{l} 
5 days \\ 
mean 
\end{tabular} 
& & Date & Confirmed & 
\begin{tabular}{l} 
New \\ 
confirmed 
\end{tabular}  
&  
\begin{tabular}{l} 
5 days \\ mean 
\end{tabular} \\ \hline
02-25 & 10     & 6    & 1,4    &  &  04-06 & 147717 & 5213 & 5676,2 \\ \hline
02-26 & 18     & 8    & 3      &  &  04-07 & 153303 & 5586 & 5337,4 \\ \hline
02-27 & 36     & 18   & 6,6    &  &  04-08 & 159051 & 5748 & 5151,4 \\ \hline
02-28 & 55     & 19   & 10,4   &  &  04-09 & 163591 & 4540 & 4951,8 \\ \hline
02-29 & 83     & 28   & 15,8   &  &  04-10 & 168151 & 4560 & 5129,4 \\ \hline
03-01 & 138    & 55   & 25,6   &  &  04-11 & 172054 & 3903 & 4867,4 \\ \hline
03-02 & 195    & 57   & 35,4   &  &  04-12 & 175087 & 3033 & 4356,8 \\ \hline
03-03 & 270    & 75   & 46,8   &  &  04-13 & 178224 & 3137 & 3834,6 \\ \hline
03-04 & 352    & 82   & 59,4   &  &  04-14 & 182662 & 4438 & 3814,2 \\ \hline
03-05 & 535    & 183  & 90,4   &  &  04-15 & 186484 & 3822 & 3666,6 \\ \hline
03-06 & 769    & 234  & 126,2  &  &  04-16 & 190308 & 3824 & 3650,8 \\ \hline
03-07 & 1101   & 332  & 181,2  &  &  04-17 & 194150 & 3842 & 3812,6 \\ \hline
03-08 & 1536   & 435  & 253,2  &  &  04-18 & 193437 & 713  & 3042,6 \\ \hline
03-09 & 2309   & 773  & 391,4  &  &  04-19 & 195655 & 2218 & 2598,6 \\ \hline
03-10 & 3285   & 976  & 550    &  &  04-20 & 198614 & 2959 & 2426   \\ \hline
03-11 & 4442   & 1157 & 734,6  &  &  04-21 & 200968 & 2354 & 2132   \\ \hline
03-12 & 5976   & 1534 & 975    &  &  04-22 & 203888 & 2920 & 1947,6 \\ \hline
03-13 & 7659   & 1683 & 1224,6 &  &  04-23 & 206002 & 2114 & 2513   \\ \hline
03-14 & 9806   & 2147 & 1499,4 &  &  04-24 & 208507 & 2505 & 2570,4 \\ \hline
03-15 & 11515  & 1709 & 1646   &  &  04-25 & 210148 & 1641 & 2306,8 \\ \hline
03-16 & 14018  & 2503 & 1915,2 &  &  04-26 & 211807 & 1659 & 2167,8 \\ \hline
03-17 & 17713  & 3695 & 2347,4 &  &  04-27 & 213338 & 1531 & 1890   \\ \hline
03-18 & 21764  & 4051 & 2821   &  &  04-28 & 214215 & 877  & 1642,6 \\ \hline
03-19 & 26333  & 4569 & 3305,4 &  &  04-29 & 215470 & 1255 & 1392,6 \\ \hline
03-20 & 31779  & 5446 & 4052,8 &  &  04-30 & 216757 & 1287 & 1321,8 \\ \hline
03-21 & 36645  & 4866 & 4525,4 &  &  05-01 & 217992 & 1235 & 1237   \\ \hline
03-22 & 41291  & 4646 & 4715,6 &  &  05-02 & 218894 & 902  & 1111,2 \\ \hline
03-23 & 48984  & 7693 & 5444   &  &  05-03 & 219338 & 444  & 1024,6 \\ \hline
03-24 & 57546  & 8562 & 6242,6 &  &  05-04 & 220362 & 1024 & 978,4  \\ \hline
03-25 & 66503  & 8957 & 6944,8 &  &  05-05 & 221236 & 874  & 895,8  \\ \hline
03-26 & 75691  & 9188 & 7809,2 &  &  05-06 & 222145 & 909  & 830,6  \\ \hline
03-27 & 83944  & 8253 & 8530,6 &  &  05-07 & 223305 & 1160 & 882,2  \\ \hline
03-28 & 90371  & 6427 & 8277,4 &  &  05-08 & 224048 & 743  & 942    \\ \hline
03-29 & 96184  & 5813 & 7727,6 &  &  05-09 & 224755 & 707  & 878,6  \\ \hline
03-30 & 104332 & 8148 & 7565,8 &  &  05-10 & 227659 & 2904 & 1284,6 \\ \hline
03-31 & 111745 & 7413 & 7210,8 &  &  05-11 & 228373 & 714  & 1245,6 \\ \hline
04-01 & 119336 & 7591 & 7078,4 &  &  05-12 & 228978 & 605  & 1134,6 \\ \hline
04-02 & 126616 & 7280 & 7249   &  &  05-13 & 229471 & 493  & 1084,6 \\ \hline
04-03 & 133294 & 6678 & 7422   &  &  05-14 & 230228 & 757  & 1094,6 \\ \hline
04-04 & 138832 & 5538 & 6900   &  &  05-15 & 230929 & 701  & 654    \\ \hline
04-05 & 142504 & 3672 & 6151,8 &  &  05-16 & 231350 & 421  & 595,4  \\ \hline
\end{tabular}
\caption{Data of the Kingdom of Spain. The list of 82 days includes the cumulative, 
new infected and mean of the previous $5$ days.}
\end{table}

\begin{figure}[!htb]
\centering \label{figure:espanha}
\includegraphics[scale=0.4]{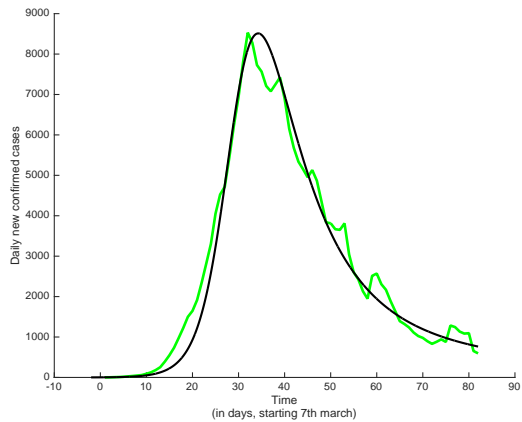}
\caption{Number of confirmed cases per day in Spain. The green line corresponds 
to the real data given Table \ref{table:spain} while the black line ($I + P + H$) 
has been obtained by solving numerically the system of fractional differential 
equations \eqref{model}, by using \cite{DIETHELM}.}
\end{figure}

By considering again the fractional order $\alpha=0.85$ and the same values of the parameters 
as in \cite{FANT}, the results of the numerical simulation are shown in Figure~\ref{figure:espanha}. 
The green line denotes the real data while the black line is the numerical solution of the 
fractional system \eqref{model}, with $N=47,000,000/425$ since in some parts of Spain 
there is more concentrated population and intensive use of public transportation.


\subsection{The Case Study of Portugal}

As for the Republic of Portugal, the data of 56 days starting 
3rd March for the cumulative cases, the new daily infected individuals, 
as well as the mean of the 5 previous days is collected in Table~\ref{table:portugal}.
\begin{table}[ht!]
\begin{tabular}{|l|l|l|l|l|l|l|l|l|} \hline
Date & Confirmed & 
\begin{tabular}{l} New \\ confirmed 
\end{tabular} 
& 
\begin{tabular}{l} 5 days \\ mean 
\end{tabular} 
& & Date & Confirmed & 
\begin{tabular}{l} 
New \\ confirmed 
\end{tabular}  
&  \begin{tabular}{l} 
5 days \\ mean 
\end{tabular} \\ \hline
03-03  & 4    & 2   & 4     &  & 03-31 & 7443  & 1035 & 725,9 \\ \hline
03-04  & 6    & 2   & 2     &  &04-01  & 8251  & 808  & 750,9 \\ \hline
03-05  & 9    & 3   & 3     &  & 04-02  & 9034  & 783  & 784,3 \\ \hline
03-06  & 13   & 4   & 4     &  & 04-03  & 9886  & 852  & 802,6 \\ \hline
03-07  & 21   & 8   & 8     &  & 04-04  & 10524 & 638  & 764,9 \\ \hline
03-08  & 30   & 9   & 9     &  & 04-05  & 11278 & 754  & 759,4 \\ \hline
03-09  & 39   & 9   & 5,5   &  & 04-06  & 11730 & 452  & 760,3 \\ \hline
03-10 & 41   & 2   & 5,3   &  & 04-07  & 12442 & 712  & 714,1 \\ \hline
03-11 & 59   & 18  & 7,6   &  & 04-08  & 13141 & 699  & 698,6 \\ \hline
03-12 & 78   & 19  & 9,9   &  & 04-09  & 13956 & 815  & 703,1 \\ \hline
03-13 & 112  & 34  & 14,1  &  & 04-10 & 15472 & 1516 & 798   \\ \hline
03-14 & 169  & 57  & 21,1  &  & 04-11 & 15987 & 515  & 780,4 \\ \hline
03-15 & 245  & 76  & 30,7  &  & 04-12 & 16585 & 598  & 758,1 \\ \hline
03-16 & 331  & 86  & 41,7  &  & 04-13 & 16934 & 349  & 743,4 \\ \hline
03-17 & 448  & 117 & 58,1  &  & 04-14 & 17448 & 514  & 715,1 \\ \hline
03-18 & 642  & 194 & 83,3  &  & 04-15 & 18091 & 643  & 707,1 \\ \hline
03-19 & 785  & 143 & 101   &  & 04-16 & 18841 & 750  & 697,9 \\ \hline
03-20 & 1020 & 235 & 129,7 &  & 04-17 & 19022 & 181  & 507,1 \\ \hline
03-21 & 1280 & 260 & 158,7 &  & 04-18 & 20206 & 1184 & 602,7 \\ \hline
03-22 & 1600 & 320 & 193,6 &  & 04-19 & 20863 & 657  & 611,1 \\ \hline
03-23 & 2060 & 460 & 247   &  & 04-20 & 21379 & 516  & 635   \\ \hline
03-24 & 2362 & 302 & 273,4 &  & 04-21 & 21982 & 603  & 647,7 \\ \hline
03-25 & 2995 & 633 & 336,1 &  & 04-22 & 22353 & 371  & 608,9 \\ \hline
03-26 & 3544 & 549 & 394,1 &  & 04-23 & 22797 & 444  & 565,1 \\ \hline
03-27 & 4268 & 724 & 464   &  & 04-24 & 23392 & 595  & 624,3 \\ \hline
03-28 & 5170 & 902 & 555,7 &  & 04-25 & 23864 & 472  & 522,6 \\ \hline
03-29 & 5962 & 792 & 623,1 &  & 04-26 & 24027 & 163  & 452   \\ \hline
03-30 & 6408 & 446 & 621,1 &  & 04-27 & 24322 & 295  & 420,4 \\ \hline
\end{tabular} 
\label{table:portugal}
\caption{Data of the Republic of Portugal. The list of 56 days includes 
the cumulative, new infected and mean of the previous $5$ days.}
\end{table}

\begin{figure}[!htb]
\centering \label{figure:portugal}
\includegraphics[scale=0.4]{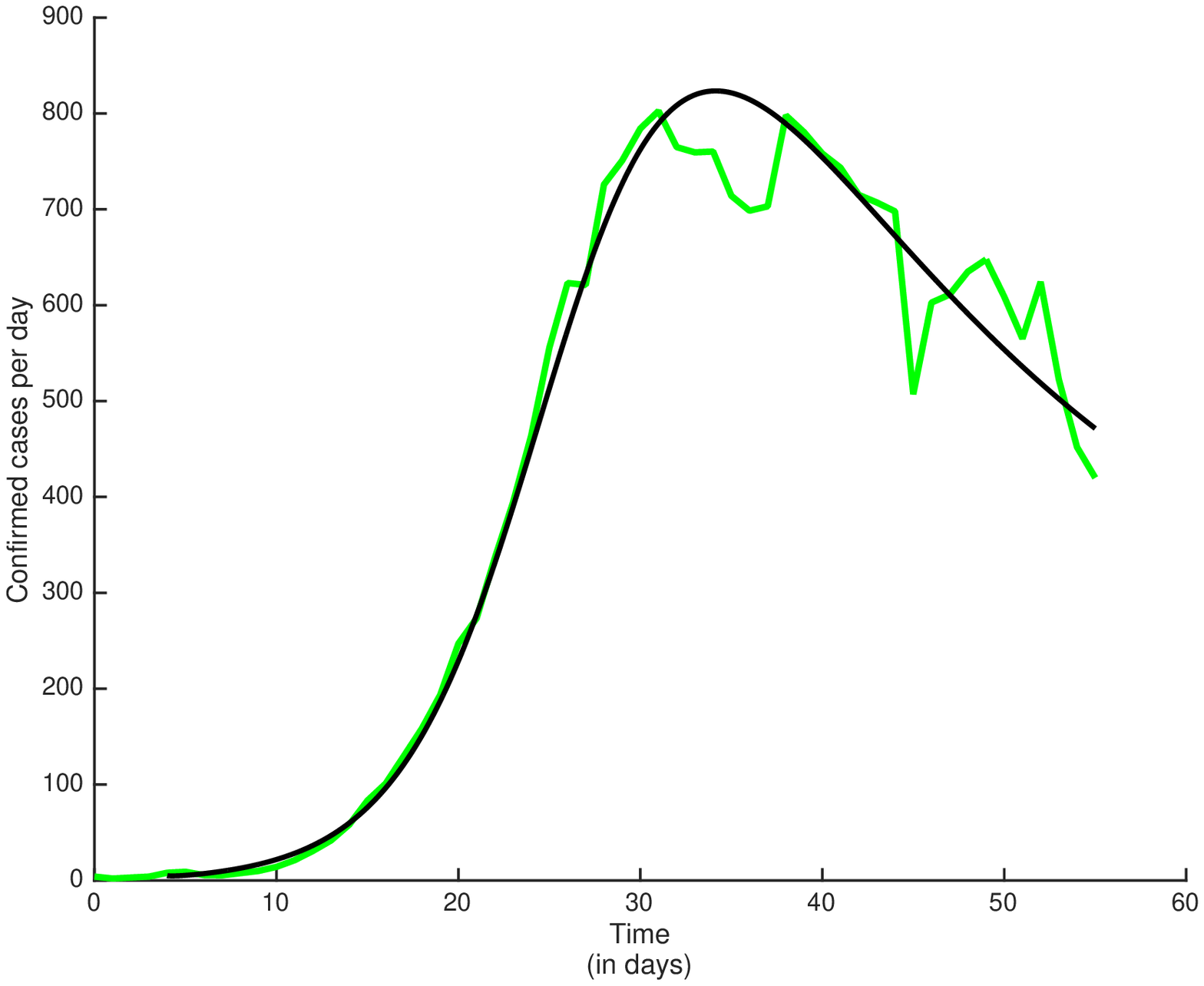}
\caption{Number of confirmed cases per day in Portugal. The green line corresponds 
to the real data given in Table~\ref{table:portugal} while the black line ($I + P + H$) 
has been obtained by solving numerically the system of fractional differential 
equations \eqref{model}, by using \cite{DIETHELM}.}
\end{figure}

By considering now the fractional order $\alpha=0.75$ and the same values of the parameters 
as in \cite{FANT}, the results of the numerical simulation are shown in Figure~\ref{figure:portugal}. 
As in the previous figures, the green line denotes the real data while the black line is the numerical 
solution of the fractional system \eqref{model}, with $N=10,280,000/1750$ since the Portuguese 
population is widely dispersed and the confinement started at an earlier stage of the spread of the disease.

\section{Conclusions and Discussion}
\label{sec:conc}

In this paper, we have shown the importance of considering a fractional Caputo 
differential system, where the order of the derivative $\alpha$ plays a crucial 
role to fit the number of confirmed cases in the regions of Galicia, Spain and Portugal.  
In fact, the considered values of  $\alpha = 0.85$ for Galicia and Spain and $\alpha = 0.75$ 
for Portugal, are not close to 1 (the classical derivative), as it happens in many 
of the proposed fractional compartmental models in the literature. Note that the same 
values of the parameters in the differential system \eqref{model}, taken from \cite{FANT}, 
were used for the three regions. Therefore, we may conclude that model \eqref{model} 
can be used to approximate the confirmed cases of COVID-19 in regions with different economic, 
geographical, social and epidemic characteristics, as it happens for the three 
considered regions in this paper. 

Our variables are divided into eight epidemiological sub-populations as in \cite{FANT} 
and they are detailed at the beginning of the second section, dedicated to the introduction 
of the dynamical model. We have solved our fractional dynamical model using a subroutine 
called \texttt{FracPECE} \cite{DIETHELM} to approximate numerically the solution of the 
proposed fractional system of differential equations. Our numerical simulations show 
a good agreement between the output of the fractional model given by the sum of the 
symptomatic and infectious individuals, super-spreaders, and hospitalized individuals 
and the data collected from the health authorities in Spain, Portugal and Galicia. 
We plan to consider other countries and regions in our future studies and also, 
of course, an update of the data. In the future, we also plan to study the stability 
of the possible equilibrium point, the bifurcation of solutions depending on the parameters, 
and the role of the basic reproduction number.

Our fractional model is novel and in the future we will study the optimal fractional 
order of differentiation for the study of the COVID-19 epidemic in different contexts. 
The system has a unique solution for given initial conditions and a detailed mathematical 
analysis study will be performed. A crucial point is, of course, to determine the optimal
fractional order $\alpha$ adequate for each process and, in this case, each region.

The results obtained here allow us to conjecture that the strains and genomes 
of the new coronavirus present in Spain and Portugal are different than those that initially hit China:
the proposed mathematical model is good to describe the outbreak that was first identified in Wuhan
in December 2019 with $\alpha =1$; to describe the spread in Spain and its autonomous community of Galicia, 
where the virus was first confirmed on January 31 and March 4 2020, respectively, with $\alpha = 0.85$;
and the COVID-19 situation in Portugal with $\alpha = 0.75$, where the first cases of COVID-19 
were recorded in March 2, 2020. We will continue our research using this and other future models, 
as well as considering different approaches as the COVID-19 evolves and new insights and conjectures emerge.


\section*{Funding}

This research was partially supported by the Portuguese Foundation for Science and Technology (FCT) 
within ``Project n.~147 -- Controlo \'Otimo e Modela\c{c}\~ao Matem\'atica da Pandemia \text{COVID-19}: 
contributos para uma estrat\'egia sist\'emica de interven\c{c}\~ao em sa\'ude na comunidade'', 
in the scope of the ``RESEARCH 4 COVID-19'' call financed by FCT; 
and by the Instituto de Salud Carlos III, within the Project COV20/00617 
``Predicci\'on din\'amica de escenarios de afectaci\'on por \text{COVID-19} a corto y medio plazo (PREDICO)'', 
in the scope of the ``Fondo COVID'' financed by the Ministerio de Ciencia e Innovaci\'on of Spain.
The work of Nda\"{\i}rou, Silva and Torres was also partially supported within project 
UIDB/04106/2020 (CIDMA); the work of Area and Nieto has been partially supported by the 
Agencia Estatal de Investigaci\'on (AEI) of Spain, cofinanced 
by the European Fund for Regional Development (FEDER) corresponding 
to the 2014-2020 multiyear financial framework, project MTM2016-75140-P.  
Moreover, Nda\"{\i}rou is also grateful to the support 
of FCT through the Ph.D. fellowship PD/BD/150273/2019;
Nieto also thanks partial financial support 
by Xunta de Galicia under grant ED431C 2019/02. Silva is also supported by national funds (OE), 
through FCT, I.P., in the scope of the framework contract foreseen in the numbers 
4, 5 and 6 of the article 23, of the Decree-Law 57/2016,
of August 29, changed by Law 57/2017, of July 19.


\section*{Acknowledgment}

The authors are grateful to the anonymous reviewers for their suggestions 
and invaluable comments.



\end{document}